\documentclass[12pt,a4paper]{article}
\usepackage{amsmath}
\usepackage{amssymb}
\usepackage{graphicx}
\usepackage{subfigure}
\usepackage{cite}
\usepackage{authblk}
\usepackage{diagbox} 
\usepackage{hyperref}
\usepackage[top=2cm, bottom=2cm, left=2cm, right=2cm]{geometry}

\begin{document}

\bibliographystyle{unsrt}

\title{Can we expect an excess of cosmological neutrinos during detection of gravitational waves?}
\author[1]{Zijian Song \thanks{zijian.song@stonybrook.edu}}
\author[1]{Xue-Qian Li \thanks{lixq@nankai.edu.cn}}
\affil[1]{School of Physics, Nankai University, Tianjin, 300071, P.R.China}
\date{} 

\maketitle

\abstract{As is well recognized, the supernova explosions produce a great amount of neutrinos, thus we have a strong reason to believe that all violent cosmological events, such as supernova explosions, gamma bursts, black hole merging or neutron star merging would cause remarkable  neutrino sprays into the open space. Recently, a very high-energy neutrino captured by the IceCube detector is believed to be radiated from a blazar. However, as the LIGO collaboration first observed gravitational wave caused by spiral approach and merging of two giant back holes, the IceCube did not see an excess of cosmological neutrinos. The reason may be due to that the source is too far away or the neutrino spectrum is not suitable for the IceCube facility detection or just missing by chance. In this work, following Hawking's picture, we suppose that neutrinos would be ejected from black holes as strong gravitational effects lower the potential barriers to enhance neutrino escape rate through quantum tunneling effects. We use the Dirac equation in the curved spacetime to describe the neutrino status and estimate the rejection rate. Then, we also adopt a simplified version where the effects of curved spacetime is ignored to clarify the physical picture. Our conclusion is drawn based on the numerical results and a discussion on the phenomenological consequence is presented. }

\begin{section}{Introduction}

\quad \ \  As everybody knows, violent cosmological events produce tremendous amount of neutrinos, such as 1987 supernova explosion \cite{bionta1991observation}, gamma bursts \cite{waxman1997high} and neutrino star merger \cite{sekiguchi2011gravitational}. Especially, a neutrino of very high energy (about 300 TeV) captured by the Ice-Cube detector is believed to be radiated from the TXS 0506+056 blazar where violent astronomical activities produce neutrino ejections in company to cosmological rays containing other particles, such as photons \cite{1807.04748} et al. This new observation stimulates a new passion for the study of the multi-messenger astronomy. Namely, researcher could investigate many signals
which may help to reveal the governing physical mechanisms and the structures of the astronomical objects. Definitely, among all the signals neutrino is the most favorable one because it can traverse a vast space without damping.

We do have a strong reason to believe that black hole merger would cause remarkable neutrino sprays into the open space \cite{dvali2016small}. The gravitational wave event GW150914 \cite{abbott2016observation} gave us a great opportunity to detect such neutrinos, but unfortunately, there was no neutrino detected by the IceCube as either temporal or spatial coincidence with the gravitational wave event \cite{adrian2016high}. To figure out the the reason how such neutrino escape detection, we should study the mechanisms which cause the neutrino radiation during the event. To make a more reasonable conclusion, a qualitative calculation is absolutely needed. Based on the model built by Hawking radiation \cite{hawking1975particle} and quantum field theory in curved spacetime \cite{birrell1984quantum, poisson2011motion, rarita1941theory}, we will try to explain the possible reason why the IceCube collaboration didn't detect such neutrino.

\par Since the quantum thermal radiance of black holes was discovered by Hawking \cite{hawking1975particle, hartle1976path, gibbons1977cosmological}, a lot of work came out concerning the physical interpretation of this phenomenon. Based on the early work of Kraus and Wilczek \cite{kraus1994some, kraus1995self, kraus1995effect}, the tunneling model which used the semi-classical WKB approximation method to describe the Hawking radiation was proposed \cite{parikh2000hawking, parikh2005energy}. By using the WKB method and complex path integration, the wave equation of the emission is obtained, and the Hawking temperature and the surface gravity is understood. The tunneling model provides a new perspective to comprehend the physical mechanism of the Hawking radiation. Furthermore, because of the striking simplicity of this model, the calculation can be generalized to particle emission in various background spacetime, such as AdS black holes \cite{hemming2001hawking},  de Sitter horizons \cite{parikh2002new, medved2002radiation}, BTZ black hole \cite{wu2006remarks, liu2006new, angheben2005hawking}, Kerr-Newman \cite{jiang2006hawking}, black rings \cite{liu2007tunneling}, Taub-NUT \cite{kerner2006tunnelling}, Vaidya \cite{jun2006tunnelling}, dynamical black hole \cite{di2007hawking}, Kerr-Godel \cite{kerner2007tunnelling}, weakly isolated horizons \cite{wu2007tunneling} and constant curvature black hole \cite{yale2011thermodynamics}. The tunneling model can also be used to study the emission of different particles such as scalars \cite{yale2011exact}, fermions \cite{kerner2008fermions, yale2011exact, chen2008hawking, chen2008hawking2, li2008dirac} and even SUSY particles, such as gravitinos \cite{yale2009gravitinos}. Recently this technique was also employed in studying quantum tunneling effects in three-dimensional spacetime \cite{ejaz2013quantum}.

\par The Intermediate binary black hole (IBBH) problem was firstly discussed by Brady, Creighton and Thorne \cite{brady1998computing}. In that paper they used a special strategy that one can change the asymptotic inertial coordinates to the spatial coordinates which co-rotate with the system. so they could carry out a numerical evolution for the binary system through the IBBH phase. Based on the work of Manasse \cite{manasse1963distortion} and D' Eath \cite{d1975dynamics, d1975interaction}, by using the Post-Newtonian method \cite{blanchet1998gravitational, will1996gravitational}, Alvi calculated the approximate metric of a binary black hole system in a circular orbit \cite{alvi2000approximate}, which makes it possible to discuss the Hawking radiation of BBH systems.

\par In this paper, following the physical picture of Hawking \cite{hawking1975particle} and Wilczek \cite{parikh2000hawking}, by using the technique of quantum field theory in curved spacetime \cite{birrell1984quantum, poisson2011motion, rarita1941theory}, we have obtained an approximative emission rate of neutrinos from the binary black hole spacetime \cite{alvi2000approximate}. Being inspired by the idea of Deeg \cite{harvey1992quantum}, Mukhanov, Wipf and Zelnikov \cite{mukhanov19944d}, we would phenomenologically provide a simplified version to estimate the rate at the quantum mechanics level as well. In this picture the event horizon is described as a potential barrier. Based on the numerical results, we would briefly discuss  possible reasons why the neutrino candidates escape detection of the IceCube Collaboration.

\end{section}

\begin{section}{Tunneling of Fermions in Curved Spacetime}

\quad \ \  First, let's consider the fermion emission from a generic metric $g^{\mu \nu}$ in the spacetime. Following Kerner,  Mann \&  Yale's work \cite{kerner2008fermions, yale2011exact}, we can have an analogous discussion on this metric. We only consider the spin-up case in this paper (The calculation or spin-down particle is similar).
\par The Dirac function in the curved spacetime is \footnote{In this paper we'll use the geometrized unit system $c=G=1$, where mass and length have the same dimension, so $\hbar$ is preserved in the equation.} \[ \left(i \gamma^{\mu} D_{\mu} + \frac{m}{\hbar}\right) \Psi(t,r,\theta,\phi) = 0,\eqno(2.1)\] in which
\begin{align}
D_{\mu} &= \partial_{\mu} + \frac{i}{2} \Gamma^{\alpha}_{\mu \beta} g^{\beta \delta} \Sigma_{\alpha \delta} \tag{2.2}\\
\Sigma_{\alpha \beta} &= \frac{i}{4} \left[ \gamma_{\alpha}, \gamma_{\beta} \right]. \tag{2.3}
\end{align}
The $\gamma$ matrices satisfy Clifford algebra, $$\left\{ \gamma_{\alpha}, \gamma_{\beta} \right\} = 2 g_{\alpha \beta} \mathbb{I}_4. \eqno(2.4)$$

\par In the curved spacetime, the metric is spacetime dependent, thus  from the Clifford algebra  the $\gamma$ matrices are also spacetime dependent. We assume in such spacetime the Dirac matrices can be written as the multiplication of the flat spacetime $\gamma$ matrices and an overall spacetime-dependent coefficient, which can be written as
$$\gamma^{t} = c_{0}(t,r,\theta,\phi) \left(\begin{array}{cc}1 & 0 \\0 & 1 \\\end{array}\right),
\gamma^{r}= c_{1}(t,r,\theta,\phi) \left(\begin{array}{cc}0 & \sigma^{3} \\\sigma^{3} & 0 \\\end{array}\right), \eqno(2.5)$$
$$\gamma^{\theta}= c_{2}(t,r,\theta,\phi) \left(\begin{array}{cc}0 & \sigma^{1} \\\sigma^{1} & 0 \\\end{array}\right),
\gamma^{\phi}= c_{3}(t,r,\theta,\phi) \left(\begin{array}{cc}0 & \sigma^{2} \\\sigma^{2} & 0 \\\end{array}\right), \eqno(2.6)$$
in which $\sigma^i$ is Pauli matrix. We know from the WKB approximation method that the tunneling probability rate $\Gamma$ can be written as
$$\Gamma \sim \exp \left( -2 \ \mathrm{Im} \ I\right), \eqno(2.7)$$
in which $I$ is the classical action of the tunneling process. So for equation (2.1), we can assume the spin-up solution of the Dirac function in a curved spacetime as,
$$\Psi_{\uparrow}(t,r,\theta,\phi) = \left(\begin{array}{cccc}H(t,r,\theta,\phi)\\0\\Y(t,r,\theta,\phi)\\0 \\\end{array}\right) \mathrm{exp} \left[ -\frac{i}{\hbar} I_{\uparrow} (t,r,\theta,\phi)\right]. \eqno(2.8)$$

\par Let us substitute the solution into the Dirac function(2.1). Because the mass of neutrino is very small, we can neglect the mass term in our calculation for simplicity. By applying WKB approximation method, the leading order of $\hbar$ should be 0 (i.e. semi-classical situation), then we get the equations,
\begin{align*}
-|c_0| H \partial_{t} I_{\uparrow} + i c_1 Y \partial_r I_{\uparrow} &= 0 \tag{2.9}\\
Y(i c_2 \partial_{\theta} I_{\uparrow} - c_{3} \partial_{\phi} I_{\uparrow}) &= 0 \tag{2.10}\\
ic_1 H \partial_{r} I_{\uparrow} + |c_0| Y \partial_{t} I_{\uparrow} &= 0 \tag{2.11}\\
Y(i c_2 \partial_{\theta} I_{\uparrow} - c_{3} \partial_{\phi} I_{\uparrow}) &= 0. \tag{2.12}
\end{align*}

Assuming $I_{\uparrow}$ as $$I_{\uparrow} = - \omega t + \mathcal{F}(r) + \mathcal{J}(\theta, \phi)., \eqno(2.13)$$
the equations become,
\begin{align*}
|c_0| H \omega+ i c_1 Y \mathcal{F}'(r) &= 0 \tag{2.14}\\
Y(i c_2 \partial_{\theta} \mathcal{J} - c_{3} \partial_{\phi} \mathcal{J}) &= 0 \tag{2.15}\\
ic_1 H \mathcal{F}'(r) - |c_0| Y \omega &= 0 \tag{2.16}\\
Y(i c_2 \partial_{\theta} \mathcal{J} - c_{3} \partial_{\phi} \mathcal{J}) &= 0. \tag{2.17}
\end{align*}

One can find two solutions from these equations,
\begin{align*}
H = -i Y, \quad \mathcal{F}^{'}_{out}(r) = \frac{|c_0|}{c_1} \omega \tag{2.18}\\
H = i Y, \quad \mathcal{F}^{'}_{in}(r) = - \frac{|c_0|}{c_1} \omega, \tag{2.19}
\end{align*}
where $\mathcal{F}_{out,in}$ represent the outgoing and incoming solutions. From the WKB approximation, we know that the tunneling amplitude is $$T \sim \mathrm{exp}\left[ -2 \mathrm{Im}(\mathcal{F}_{out} - \mathcal{F}_{in} )\right], \eqno(2.20)$$
where $$\mathrm{Im} (\mathcal{F}_{out} - \mathcal{F}_{in}) = \mathrm{Im} \left( \int^{r_{H}+\epsilon}_{r_{H}-\epsilon} 2 \omega \frac{|c_0|}{c_1} \mathrm{d} r\right) = \mathrm{Im} \left( \int^{r_{H}+\epsilon}_{r_{H}-\epsilon} \frac{2 \omega}{\left( \frac{c_1}{|c_0|}\right)} \mathrm{d}r\right), \eqno(2.21)$$
where $r_{H}$ is the radius of the event horizon. From the Clifford algebra, we know that,
$$c_0 = \sqrt{g^{00}}, c_1 = \sqrt{g^{11}}. \eqno(2.22)$$

\par By now, we have got (2.20), (2.21) and (2.22) which determine the tunneling amplitude in a generic spacetime. But a point needs to clarify, although we use a generic spacetime $g^{\mu \nu}$ from the very beginning, we still cannot say those formulae work for all spacetimes, because of the assumption (2.13), which demands the spacetime to possess a spherical symmetry. But we will show later, although strong gravity will change the shapes of the two concerned black holes, the deformation is negligible. We can still calculate the tunneling amplitude as they are accounted as spherical Schwarzschild black holes and we may add a correction induced by gravity at the end of our calculation. So we will still keep the assumption (2.13) without losing generality. Actually, if we choose the coefficients as $$c_0 = \frac{i}{\sqrt{1-\frac{r_H}{r}}}, \ c_1 = \sqrt{1-\frac{r_H}{r}}, \ c_2 = \frac{1}{r}, \ c_3 = \frac{1}{r \sin \theta}, \eqno(2.23)$$ the result would accord with that in the original paper \cite{kerner2008fermions, parikh2000hawking}.

\end{section}

\begin{section}{The Metric of a Binary Black Hole System}

\quad \ \  For a binary black hole system, one black hole exerts a gravitational force on the other one, which can be called as the tidal deformation. A Schwarzschild black hole is spherically symmetric, so the radius of the event horizon $r_H$ is a constant uniquely determined by its mass. Under the effect of tidal force, the shape of event horizon becomes flatter and the radius becomes spacetime-dependent as well. From our previous discussion, we know the tunneling process happens near the event horizon, so it is crucial to calculate the shape of these black holes, i.e. a new metric for such system. The detailed calculation of the metric of a binary black hole system can be found in K. Alvi's paper \cite{alvi2000approximate}. We will only show some relevant results here.

\begin{figure}[ht]
\centering
\includegraphics[width=5in]{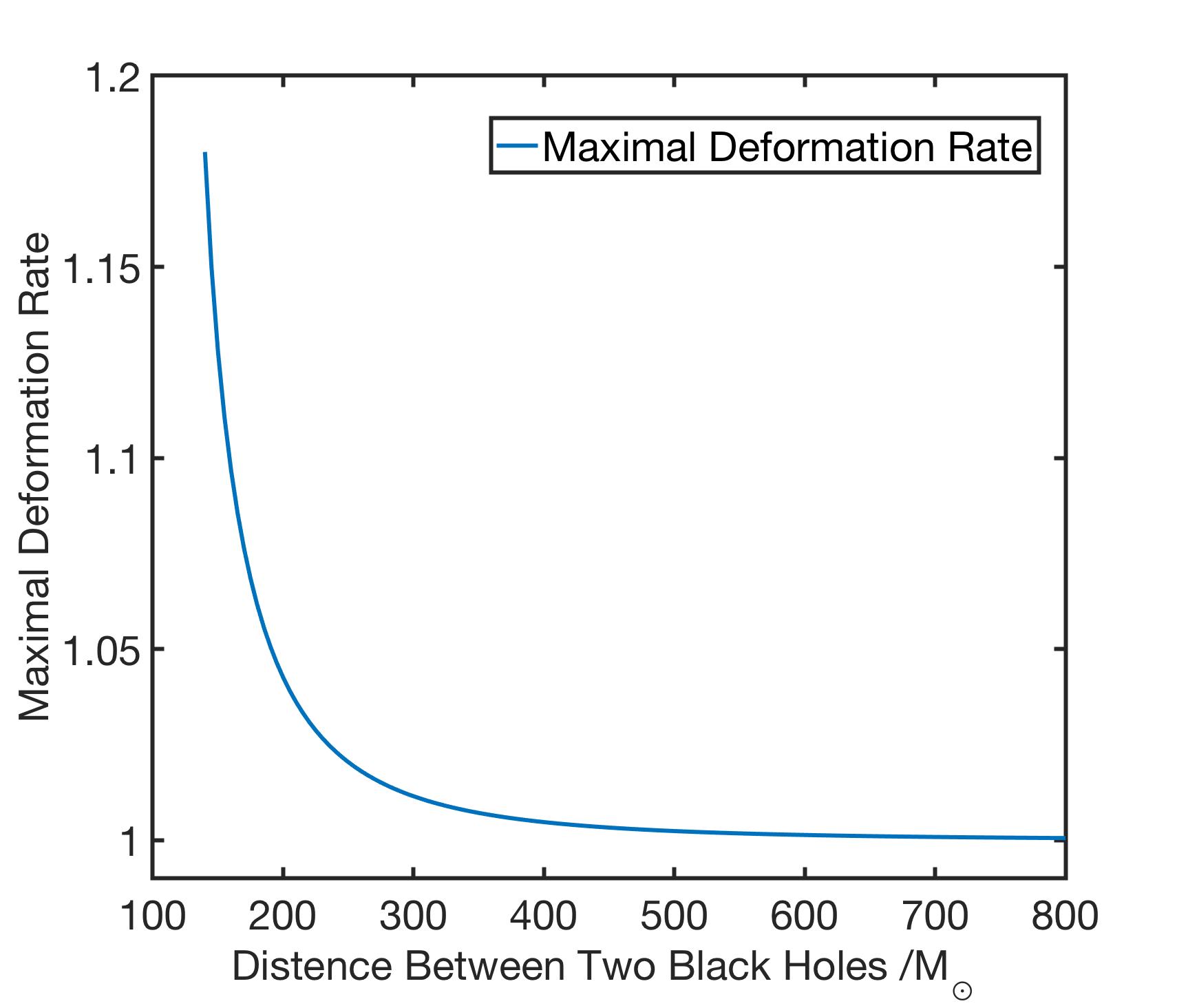}
\caption{The X-label represents the distance between two black holes in the unit of the mass of the Sun. The Y-label represents the maximal deformation rate. We find even when the two black holes are very close, the deformation rate is still not significant, which means we can approximately regard them as spherical Schwarzschild black holes.}
\label{fig:subfig}
\end{figure}

\par Near the horizon, Alvi got the result that the components of the metric are
\begin{align*}
g_{00} &= -1 + \frac{2 m_1}{R} + \frac{m_2}{b^3} \left[ 3(X \cos \Omega T + Y\sin \Omega T)^2 - R^2\right], \tag{3.1}\\
g_{ij} &= \delta_{ij} \left\{ 1 + \frac{2 m_1}{R} + \frac{m_2}{b^3} \left[ 3 (X \cos \Omega T + Y \sin \Omega T)^2 - R^2\right]\right\}. \tag{3.2}
\end{align*}
$R, \theta, \phi$ are the polar coordinates of the first black hole and $m_1, m_2$ are the masses of these two black holes. $b$ is the distance between the centers of the two black holes. $X = R \sin \theta \cos \phi$, $Y= R \sin \theta \sin \phi$, $\Omega = \omega [1 - \frac{\mu}{b} + O(\epsilon^3)]$, $\omega$ is the angular velocity of the system and $\mu$ is the reduced mass of the system. We only show the diagonal components. Other components are omitted for simplicy.

\par By solving the equation $$\frac{c_1}{|c_0|} = 0, \eqno(3.3)$$ we can get the radius of the event horizon in different directions. We define the ratio of the radius after and before deformation in the  direction with largest deformation as the maximal deformation rate. Figure 1 shows the relation between the maximal deformation rate and the distance between two black holes. In this paper, we use the gravitational wave event GW150914 as a sample. This system consists of two black holes with 36 $M_{\odot}$ and 29 $M_{\odot}$ respectively \cite{abbott2016observation}, which means the radius of each black hole is 72 $M_{\odot}$ and 58 $M_{\odot}$ in geometrized unit system. \footnote{In geometrized unit system, the Schwarzschild metric can be written as $$ds^2 = - \left( 1 - \frac{2M}{r} \right) dt^2 + \left( 1 - \frac{2M}{r} \right)^{-1} dr^2 + r^2 \left( d\theta^2 + \sin^2 \theta d\phi^2 \right),$$ which indicates the radius of the event horizon is $2M$.} We can notice from Figure 1 that the maximal deformation rate \footnote{which means the deformation rate on the direction between these two black holes.} only reaches 1.18 when the distance is $150 M_{\odot}$. So we can see that the deformation effect is not significant even though they are very close. In other direction, the rate is much smaller than the maximal case. By now we know that the assumption we have made in the previous section is valid. The shape of the black hole and the deformation rate in different direction are shown in Figure 2 and  Table 1 in the Appendix.

\begin{figure} [ht]
\centering
\subfigure[]{\label{fig:subfig:a}
\includegraphics[width=0.45\linewidth]{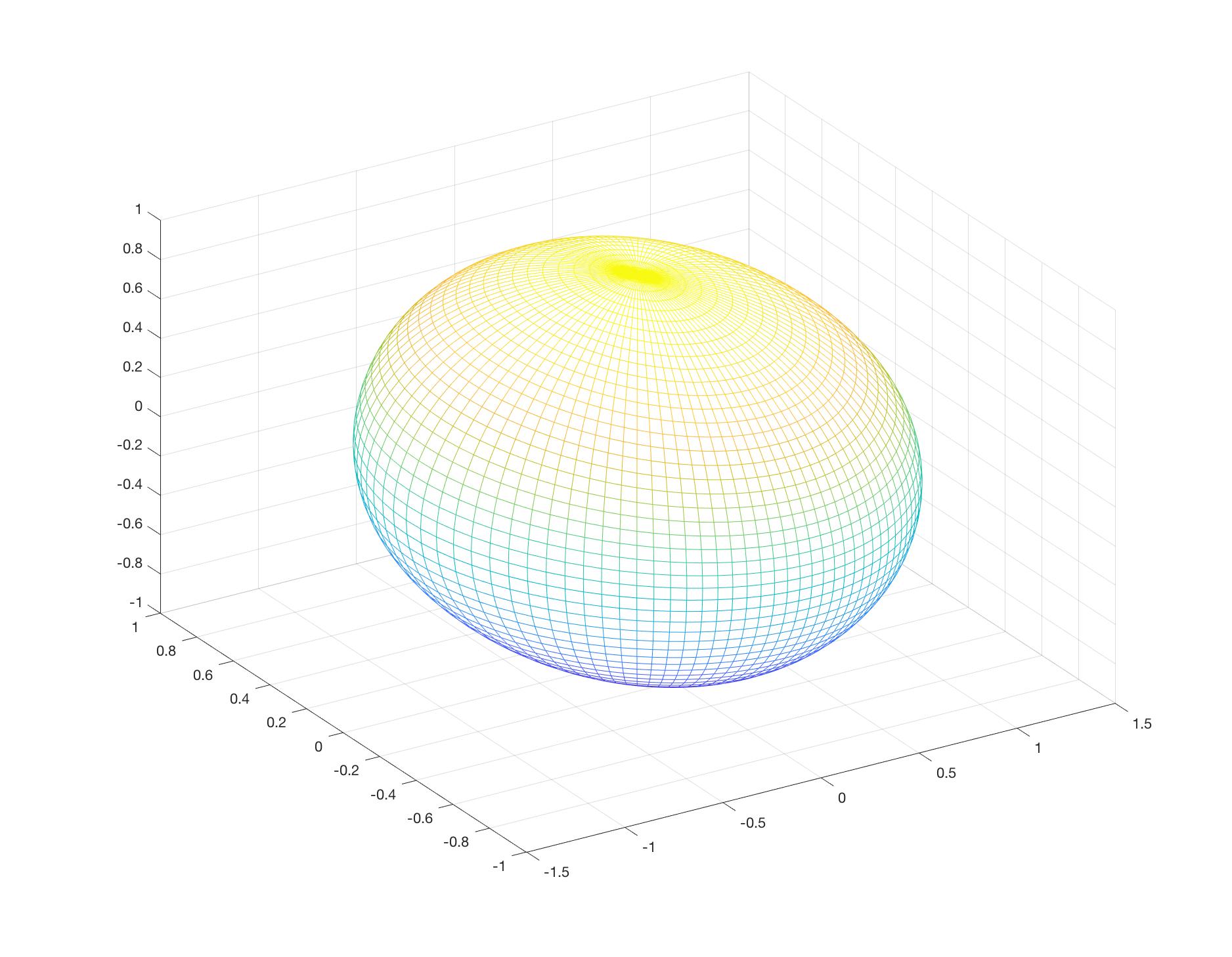}}
\hspace{0.01\linewidth}
\subfigure[]{\label{fig:subfig:b}
\includegraphics[width=0.45\linewidth]{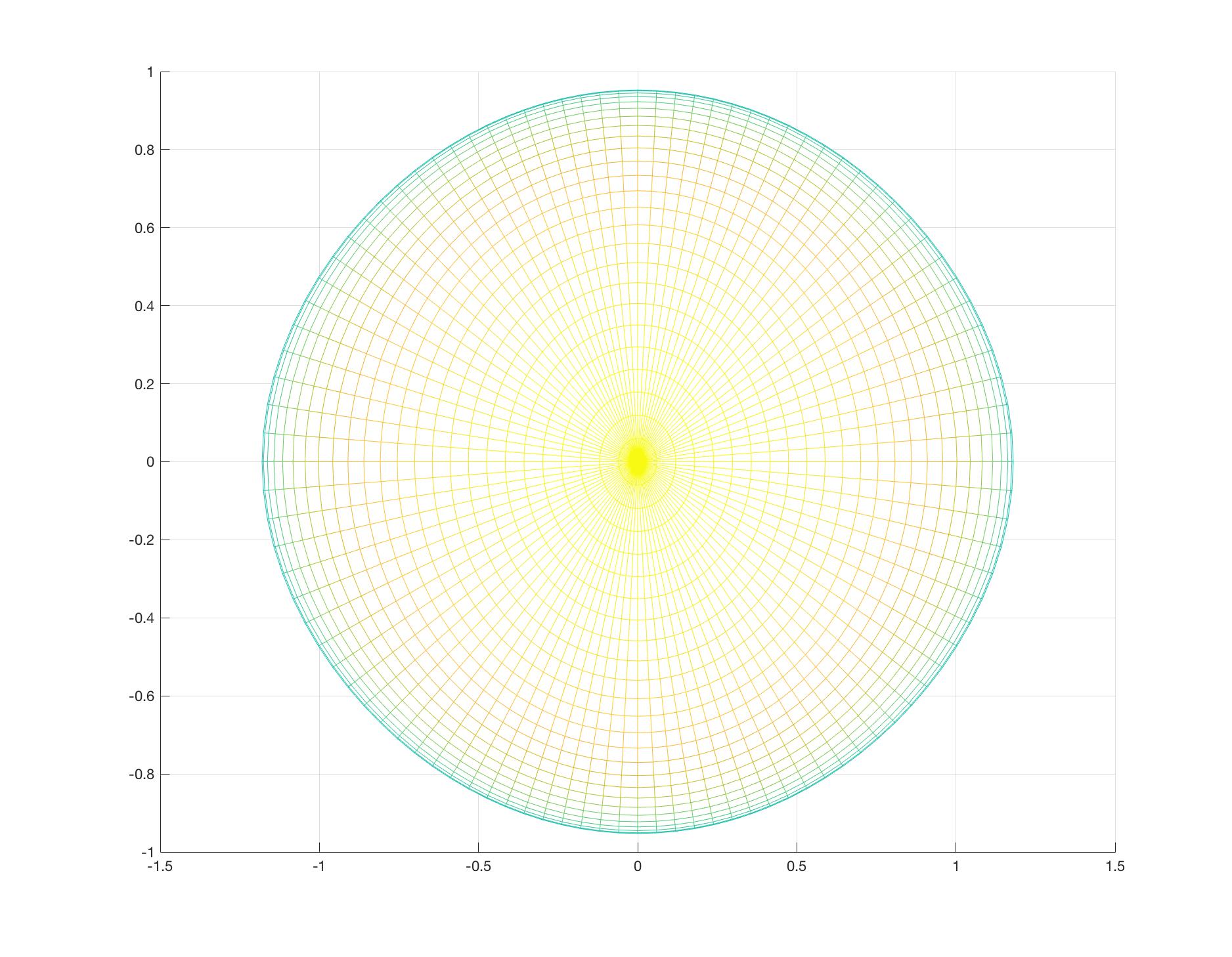}}
\caption{These two pictures show the deformation of the black hole under strong gravity. When the black holes in GW150914 $140M_{\odot}$ far away from each other, the shape of the bigger one is shown in the two pictures. We can know from (b) that the radius at the direction between two black holes increases (the X direction), and decreases at the vertical direction.}
\label{fig:subfig}
\end{figure}

\par A subtle thing we need to notice in the calculation is that the $c_i$ coefficients are defined with the contravariant tensor, such as in (2.5) and (2.6). So to make it proper to calculate in (3.1) and (3.2), we need to lower the index by $g_{\mu\nu}$.
\end{section}

\begin{section}{Tunneling from a BBH System}

\quad \ \  From the metric near the horizon of the first black hole, combining with (3.1), (3.2) and (3.3), we notice that for $R > 2 m_1$, $$\frac{c_1}{|c_0|} \approx 1 - \frac{2 m_1}{R} - \frac{m_2}{b^3} \left[ 3 (X \cos \Omega T + Y \sin \Omega T)^2 - R^2\right]. \eqno(4.1)$$ Defining $V = \frac{c_1}{|c_0|}$, we  notice that the positive solution of the equation with $V$ represents the position on the event horizon while the negative solution is abandoned due to the physical consideration. We can set $V$ in (2.21) and integrate it, and then substitute the result into (2.20). Then the tunneling amplitude can be calculated. During the calculation we find that there is a pole when $V=0$, so analytic continuation of the domain and integrate it on the complex plane is needed.

\par Near the horizon $V$ can be Taylor expanded, $$V(t,r, \theta, \phi) = V'(t,r_{H}, \theta, \phi) (r - r_{H})+.... \eqno(4.2)$$ By selecting the upper contour, the integration becomes
\begin{align*}
\mathrm{Im} (\mathcal{F}_{out} - \mathcal{F}_{in}) &= \mathrm{Im} \left( \int^{r_{H} + \epsilon}_{r_{H}- \epsilon} \frac{2\omega}{V(t,r,\theta,\phi)} \mathrm{d}r\right)\\
&= \mathrm{Im} \left( \frac{1}{V'(t,r_{H},\theta,\phi)}\int^{r_{H} + \epsilon}_{r_{H}- \epsilon} \frac{2\omega}{r - r_{H}} \mathrm{d}r\right)\\
&= \frac{\pi \omega}{\kappa}, \tag{4.3}
\end{align*}
where $\kappa = \frac{1}{2} V'(t,r_{H},\theta,\phi)$ is the surface gravity for the Hawking radiation. For a Schwarzschild black hole, the surface gravity is a function of mass. However, because of the strong gravity, the spacetime is not isotropic anymore and the surface gravity becomes spacetime-dependent as well.

\par We also use the gravitational wave event GW150914 as an example to show the difference of neutrino emission before and after occurrence of the strong gravity \cite{abbott2016observation} . The masses of the two black holes are $36^{+5}_{-4} M_{\odot}$ and $29^{+4}_{-4} M_{\odot}$, where $M_{\odot}$ is the mass of the Sun. In our theory, when two black holes approach to each other, the shape of their event horizons will change because of tidal force as it is shown in Figure 2, as aforementioned. Combining with the calculation we did in Section 2 and 3, we note that such deformation changes the surface gravity which we get in (4.3), and further changes the emission rate of neutrino. It is rather difficult to make an accurate computation on absolute values of the emission rate because of existence of some unknown overall coefficients, but a ratio of the emission rate after and before occurrence of the influence by gravity is easy to get, which is useful for experimental observation.

\begin{figure} [ht]
\centering
\subfigure[]{\label{fig:subfig:a}
\includegraphics[width=0.45\linewidth]{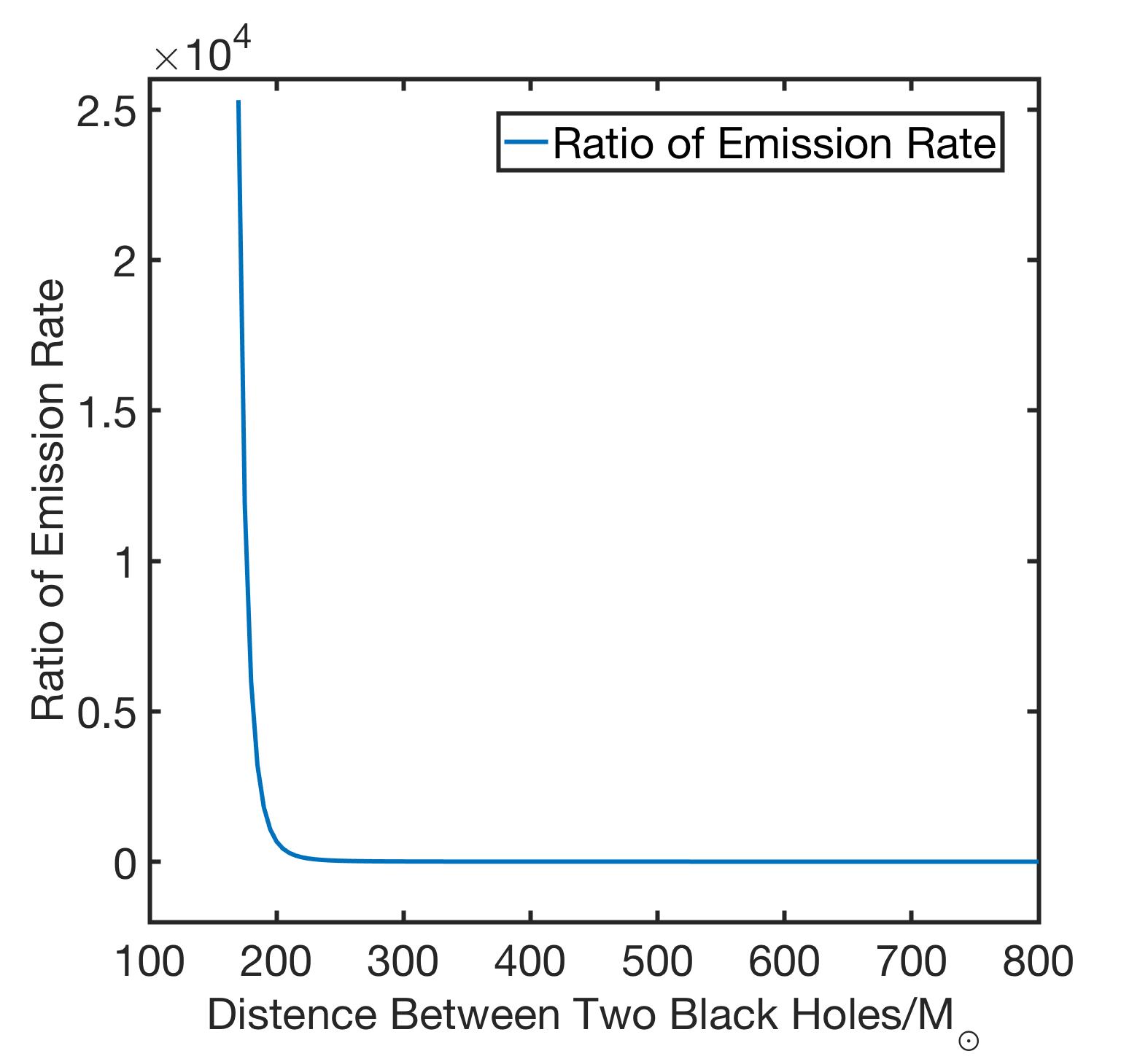}}
\hspace{0.01\linewidth}
\subfigure[]{\label{fig:subfig:b}
\includegraphics[width=0.45\linewidth]{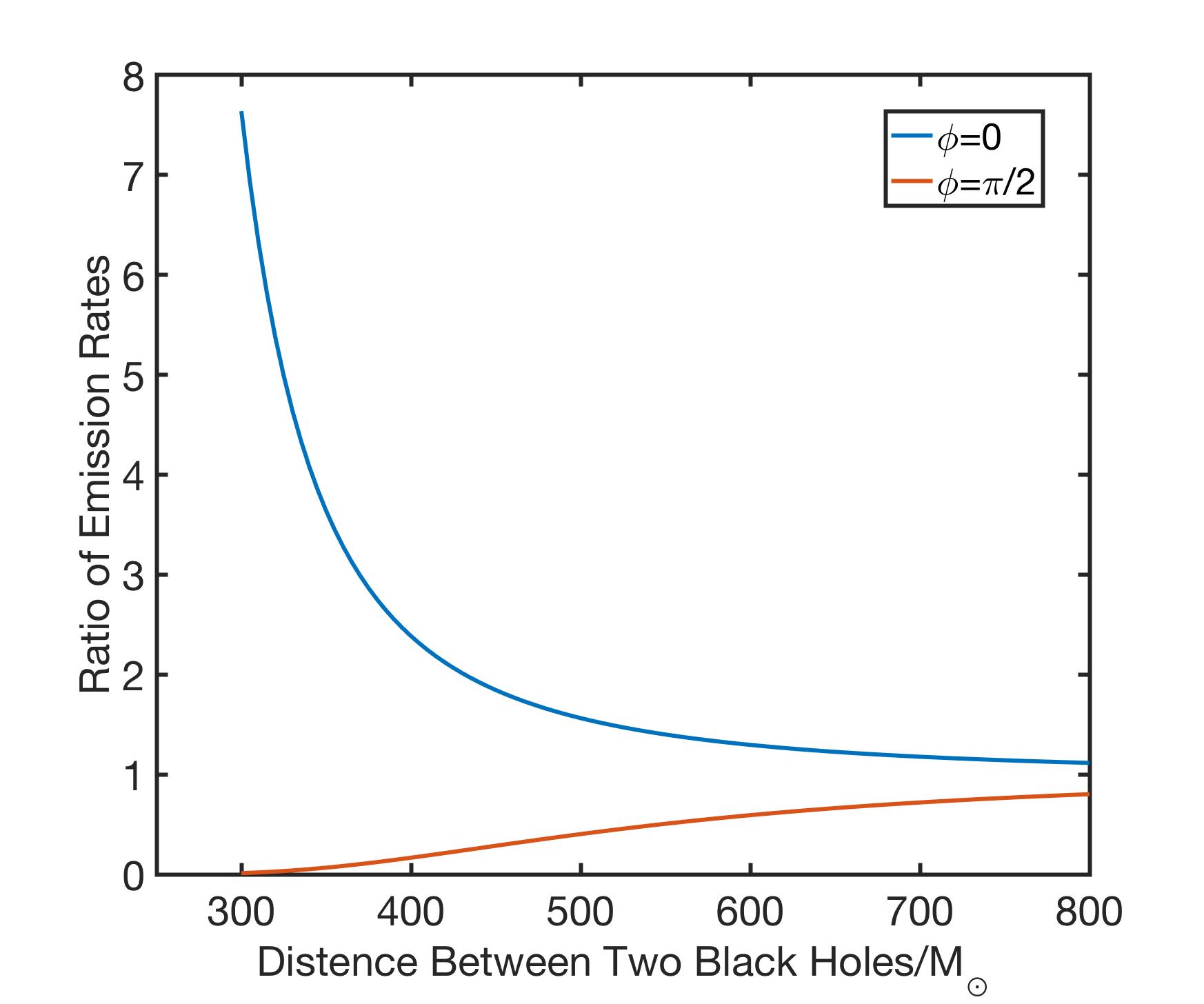}}
\caption{These two pictures show the deformation of the black hole under strong gravity. When the black holes in GW150914 $140M_{\odot}$ far away from each other, the shape of the bigger one is showed in the two pictures. We can know from (b) that the radius on the direction between two black holes increase (the X direction), and decrease on the vertical direction.}
\label{fig:subfig}
\end{figure}

\par Because this event happened at $410^{+160}_{-180} \ Mpc$ away from us,  we set the center of one of the black holes as the origin of the system. Due to the relation
$$T \sim \mathrm{exp}\left[ -2 \mathrm{Im}(\mathcal{F}_{out} - \mathcal{F}_{in} )\right], \eqno(4.4)$$ the ratio of the emission rate after and before the tidal deformation is $$\gamma = \frac{e^{-2 \pi \omega/\kappa_{1}}+e^{-2 \pi \omega/\kappa_{2}}}{e^{-8 \pi \omega m_{1}}+e^{-8 \pi \omega m_{2}}}, \eqno(4.5)$$ where $\kappa_{1}, \kappa_{2}$ is the surface gravity for each black hole. The numerator represents the emission rate when the interaction between each black hole occur and the denominator represents the emission rate before occurrence of the interaction. The ratios of the emission rate in different direction are shown in Table 2 of the Appendix.

\par Another point needs to be taken into account is how to define the azimuth of the observation to the BBH system. A simple set is to let the rotating plane be vertical to the observation direction, which means $\theta=0$ in the spherical coordinate. As the two black holes make a spiral motion and get closer to each other, the change of the ratio of emission rate is shown in Figure 3(a). It is found, when they are far away from each other, the ratio is approximately 1. But when the distance is shorter than $200 M_{\odot}$, the ratio sharply increases. It even reaches an order of $10^4$ when the distance is $170 M_{\odot}$, which definitely may cause observational effects.

\par Another special case is that the observatory is on the rotating plane, which means $\theta=\pi/2$. Because of rotation, the emission rate towards us will also periodically changes. When $\phi=0$ the maximal value appears whereas the minimal value appears at $\phi=\pi/2$. Figure 3(b) shows the upper and lower bound of the ratio of emission rate. It changes periodically between the upper and lower bounds and as the frequency increases, we can expect a chirp signal.

\par However, more generally speaking, the observatory has an arbitrary azimuth with respect to the rotating plane, so that the real bounds would reside inside that shown in Figure 3(b).

The numerical results are shown in the Appendix.

\par In conclusion, the pattens of the neutrino emission depends on the relative position between the binary black hole system and the observatory. Usually the intensity may oscillate and as the black holes approaching, the amplitude increases.

\end{section}

\begin{section}{A Semi-Classical Interpretation of the Tunneling Process}

\quad \ \ In quantum mechanics, the WKB  method is a semi-classical approximation to evaluate the tunneling rate through a potential barrier. The tunneling we discussed in previous sections is a physical effect caused by the event horizon of a black hole which is equivalent to a potential barrier. We can regard the event horizon as a potential barrier and study that problem at quantum mechanical level, which may make the picture simpler and clearer. In other words, in our case, the gravity effect increases the tunneling rate of neutrinos,
that is somehow in analog to the effects that an adjacent strong electric field lowers the potential barrier which prevents the electrons to escape from metal.

\par The tunneling amplitude can be written as $$T=\mathrm{exp} \left[ -\frac{2}{\hbar} \int^a_b \sqrt{2m (P(r)-E)} \mathrm{d} r \right], \eqno(5.1)$$ where $a, b$ are the turning points. Now we need to introduce an effective potential $P(r)$. (Since the symbol $V$ was used as components of the metric in last section, here we employ the symbol $P(r)$ for the potential).

\par For a binary black hole system, each black hole is influenced by the other one. Because of the tidal deformation, the potential also changes. As a toy model, we would simply regard the uninfluenced potential in a Gaussian form. Being perturbed by the gravity, the potential becomes $$P(r,\theta,\phi) = P(\theta, \phi) \mathrm{exp} [(r-r_H)^2] \eqno(5.2)$$
Then the tunneling amplitude in $(\theta, \phi)$ direction turn out to be $$T \sim \mathrm{exp} \left[ -\frac{2}{\hbar} \int^a_b \sqrt{2m (P(\theta, \phi) \mathrm{exp} [(r-r_H)^2]-E)} \mathrm{d} r \right]. \eqno(5.3)$$
Without losing generality, we can simply re-scale the integration to $$T' = \mathrm{exp} \left[ -\frac{2}{\hbar} \int^{r_1}_{-r_1}  \sqrt{2mE (W(\theta, \phi) \mathrm{exp} (r^2)-1)} \mathrm{d} r \right], \eqno(5.4)$$ where $r_1$ is the turning point and $W(\theta, \phi) = P(\theta, \phi) / E.$

\par From our early discussion, the tunneling amplitude in the $(\theta, \phi)$ direction is also proportional to $$T \sim \mathrm{exp} [-\frac{2}{\hbar} \frac{\pi \omega}{\kappa}]. \eqno(5.5)$$ Now we have the relation between the angular coefficient and the surface gravity,
$$\frac{1}{\kappa} \sim \int^{r_1}_{-r_1} \sqrt{W(\theta, \phi) \mathrm{exp} (r^2) - 1} \mathrm{d} r \eqno(5.6)$$

\par Based on the discussion of the Hawking radiation,  $\frac{1}{\kappa}$ is proportional to the mass of the black hole, so it is a  huge quantity. And from the analysis of the integration, we find that the result increases with the increase of $W(\theta, \phi)$. So for an actual surface gravity, the quantity of $W(\theta, \phi)$ is also huge, so that the influence of the small term in the integrand can be neglected, which means $$\int^{r_1}_{-r_1} \sqrt{W(\theta, \phi) \mathrm{exp} (r^2) - 1} \mathrm{d} r \approx \int^{r_1}_{-r_1} \sqrt{W(\theta, \phi) \mathrm{exp} (r^2)} \mathrm{d} r, \quad W(\theta, \phi) \to \infty. \eqno(5.7)$$
Then we can find the relation between $W(\theta,\phi)$ and $\kappa$,
\begin{align*}
\frac{1}{\kappa} &\sim \int^{r_1}_{-r_1} \sqrt{W(\theta, \phi) \mathrm{exp} (r^2)} \mathrm{d} r\\
&\sim \sqrt{W(\theta,\phi)} \cdot \sqrt{\frac{\pi}{2}}. \tag{5.8}
\end{align*}
So we can writte $W(\theta, \phi)$ as $\kappa^{-2}$ multiplying a overall constant,
$$W(\theta, \phi) = \frac{C}{\kappa^2}. \eqno(5.9)$$

\par By substituting this relation into the equation, $$\int^{r_1}_{-r_1} \sqrt{W(\theta, \phi) \mathrm{exp} (r^2)} \mathrm{d} r = \frac{\pi \omega}{\kappa}, \eqno(5.10)$$ we obtain the overall constant $C$ as $C=\frac{\pi \omega}{m E}$, the potential then becomes $$P(r,\theta,\phi) = \frac{\pi \omega^2}{m \kappa^2(\theta, \phi)} \mathrm{exp} [(r-r_{H})^2], \eqno(5.11)$$ in which $\theta, \phi$ are included in the surface gravity, $\omega$ is the energy of a single particle, $m$ is the inertial mass of a particle and $E$ is the energy of a particle in the black hole.

\end{section}

\begin{section}{Discussion}

\quad \ \  In this paper, we have generalized the result of Kerner and Mann \cite{kerner2008fermions} to a generic spacetime, which allows us to study the neutrino emission in a binary black hole background. We  provide an approximate estimate of the ratio of the emission from the BBH system over the background radiation, which may be useful for experimental observations. We find in figure 2 that there is a significant increase of the ratio when two black holes are getting very close. This seems to show that there should be a bump peaked up from the cosmological neutrino background radiated by various sources. Unfortunately, the IceCube Collaboration did not find the neutrino candidates accompanied with GW150914 \cite{adrian2016high}.

\par The possible reason we think is that the emission rate of neutrino of isolated black holes is not sufficiently large. The emission rate is proportional to $\exp (-8 \pi \omega M)$, where $M$ is an extremely tremendous quantity, so although the emission rate is hundreds even thousands times larger, it is still not much enough for our detectors to observe.

\par Another factor needs to be addressed is that different from the potential barrier in electrodynamics, the potential barrier existing at the horizon surface of black holes which we discussed in Section 5 is of a special property as: the larger the energy of the particles which are supposed to penetrate through the barrier is, the higher the barrier would be, which is shown in (5.11). Why can it happen? Because the effective barrier is caused by the gravity of the black hole, and the energy is equivalent to the gravitational mass, so that as the energy is larger, the gravitational attraction is stronger. Phenomenologically it implies that the effective barrier should be higher for more energetic particles.

In quantum electrodynamics, as the potential barrier is fixed, the tunneling amplitude is only decided by the kinetic energy of the particles, which means that we can get the information inside the barrier from the outgoing particles. That is obviously not the case for black holes. Namely, the spectrum of the outgoing neutrino is just like blackbody radiation, which does not bring out detailed information inside the cavity. In this picture,  the possibility of radiating high energy neutrinos is rarer than that of lower energies.
Therefore, the majority of the emitted neutrinos has lower energies which is beyond the measuring scope of the IceCube detector. That may indeed be the reason why the IceCube failed
to observe excess neutrino during the blackhole and neutron star merging process when gravitational waves were radiated. By our calculation, there are indeed neutrino spray from the black holes, but just missed out the IceCube detector.

\par However, the model we proposed is just a toy model. We only talked about the neutrino emission before merging, i.e. the spiral stage yet, because we do not have a reasonable model to predict the neutrino ejection during the merging stage. There would be more violent astronomical activities than the spiral process and the resultant black hole would be much heavier than the two parent black holes and possesses a larger horizon area. Whether there will be a remarkable ejection of neutrinos when the binary black hole system is merging is still an open problem, and the emission rate of other types of black holes should be calculated as well. We will employ a tentative model to deal with the neutrino radiation during the merging stage and probably the glow after.

\par We also provide a semi-classical scenario to predict neutrino emission rate for the spiral stage. The potential of scattering as shown is different for different energies. For high energy neutrino, the potential would be higher and vice verse. In our model we suppose there is a uniform distribution of energy for neutrinos in the black hole. Surely, we can also set different distribution of energy, which may cause more interesting results.

\par Actually, there is no any reason that a great amount of neutrinos is not ejected during processes of black holes or neutron stars merging. But the possibility of capturing them at
any earth detectors is still slim and moreover, the parent astronomical objects are too far away, the solid angle to the earth is small, thus even though the amount of neutrinos ejected during the violent activities is huge, the probability of missing them is quite large. Namely the ejected neutrinos accompanying the gravitational waves indeed existed, but just missed our detectors by chance. Therefore, we would like to encourage our experimental colleagues to continue exploring such cosmological neutrinos at the IceCube, the JinPing underground observatory of China and other facilities around the world.

\par Now let us emphasize the main points of our conclusion. Towards the results we derived in the text, actually, we can only obtain a relative value of the radiation rate with respect to the background of cosmological neutrino flux. As observed, when two black holes approach each other, i.e. at the spiral stage, the radiation rate would be enhanced by $10^4$ times or even more. However, as the background is not well determined so far, we only know the relative rate, but not the absolute value. As commonly recognized, the background of neutrino flux from pure Hawking’s radiation is too small to be measured at present. But due to the huge enhance factor, it might reach an observational level.

\par The earlier observation of absence of high energy neutrino at IceCube might hint something, but as we discussed, they might just escape detection by chance. In fact, the production rate of high energy neutrinos is greatly enhanced, even though not as fast as low energy neutrinos.
  
\par Indeed, we only can lay the hope that arrival of high energy neutrinos and/or some excess of cosmological neutrinos in accompany of the gravitational waves would be observed by our earth detector.
Therefore, as we say, the higher the energy of the outgoing neutrinos is, the more difficult the escape via tunneling would be, it is only relative to the low energy ejection, but indeed, its tunneling rate is still obviously enhanced and may be caught by the IceCube detector.

\end{section}

\begin{section}{Appendix}

\begin{table}[!htbp]
\centering
\begin{tabular}{|c|c|c|c|c|c|c|}
\hline
\diagbox{$\theta$}{$\phi$}&$0$&$2\pi/5$&$4\pi/5$&$6\pi/5$&$8\pi/5$&$2\pi$\\ 
\hline
$0$&0.9526&0.9526&0.9526&0.9526&0.9526&0.9526\\
\hline
$\pi/5$&1.0020&0.9567&0.9832&0.9832&0.9567&1.002\\
\hline
$2\pi/5$&1.1385&0.9636&1.0490&1.0490&0.9636&1.1385\\
\hline
$3\pi/5$&1.1385&0.9636&1.0490&1.0490&0.9636&1.1385\\
\hline
$4\pi/5$&1.0020&0.9567&0.9832&0.9832&0.9567&1.002\\
\hline
$\pi$&0.9526&0.9526&0.9526&0.9526&0.9526&0.9526\\
\hline
\end{tabular}

\caption{Deformation rate in different dirrections when the distance between two black holes is 140$M_{\odot}$.}

\end{table}

\begin{table}[!htbp]
\centering
\begin{tabular}{|c|c|c|c|c|c|c|}
\hline
\diagbox{$\theta$}{$\phi$}&$0$&$2\pi/5$&$4\pi/5$&$6\pi/5$&$8\pi/5$&$2\pi$\\ 
\hline
$0$&$1.7574 \times 10^7$&$1.7574 \times 10^7$&$1.7574 \times 10^7$&$1.7574 \times 10^7$&$1.7574 \times 10^7$&$1.7574 \times 10^7$\\
\hline
$\pi/5$&0.4708&$4.4465 \times 10^6$&460.3954&460.3954&$4.4465 \times 10^6$&0.4708\\
\hline
$2\pi/5$&$7.6296 \times 10^{-27}$&$4.3150 \times 10^{5}$&$3.5626 \times 10^{-9}$&$3.5626 \times 10^{-9}$&$4.3150 \times 10^{5}$&$7.6296 \times 10^{-27}$\\
\hline
$3\pi/5$&$7.6296 \times 10^{-27}$&$4.3150 \times 10^{5}$&$3.5626 \times 10^{-9}$&$3.5626 \times 10^{-9}$&$4.3150 \times 10^{5}$&$7.6296 \times 10^{-27}$\\
\hline
$4\pi/5$&0.4708&$4.4465 \times 10^6$&460.3954&460.3954&$4.4465 \times 10^6$&0.4708\\
\hline
$\pi$&$1.7574 \times 10^7$&$1.7574 \times 10^7$&$1.7574 \times 10^7$&$1.7574 \times 10^7$&$1.7574 \times 10^7$&$1.7574 \times 10^7$\\
\hline
\end{tabular}

\caption{The ratio of emission rate in different dirrections when the distance between two black holes is 140$M_{\odot}$.}

\end{table}

\end{section}

\bibliography{blackhole}

\begin{section}{Acknowledgement}

This work is supported by the National Natural Science Foundation of China under contract No.11075079, 11135009 and 11005061.

\end{section}

\end{document}